\RequirePackage{fix-cm}
\documentclass[twocolumn,epjc3,showpacs,preprintnumbers,amsmath,amssymb]{svjour3}
\smartqed  
\RequirePackage{graphicx}
\RequirePackage[numbers,sort&compress]{natbib}
\usepackage{graphicx,amsmath,amsfonts,latexsym,amssymb,graphics,epsfig,subfigure,color,makeidx,hyperref}
\usepackage{multirow}

\journalname{Eur. Phys. J. C}

\newcommand\beq{\begin{equation}}
\newcommand\eeq{\end{equation}}
\newcommand\beqn{\begin{eqnarray}}
\newcommand\eeqn{\end{eqnarray}}
\newcommand\bal{\begin{align}}
\newcommand\eal{\end{align}}
\newcommand\nn{\nonumber}
\newcommand\fc{\frac}
\newcommand\lt{\left}
\newcommand\rt{\right}
\newcommand\pt{\partial}

\begin{document}

\allowdisplaybreaks

\title{Born-Infeld Black Holes in 4D Einstein-Gauss-Bonnet Gravity}

\author{Ke Yang\thanksref{e1,addr1}
       \and Bao-Min Gu\thanksref{e2,addr2,addr3}
       \and Shao-Wen Wei\thanksref{e3,addr4,addr5}
       \and  Yu-Xiao Liu\thanksref{e4,addr4,addr5}
}

\thankstext{e1}{e-mail: keyang@swu.edu.cn}
\thankstext{e2}{e-mail: gubm@ncu.edu.cn}
\thankstext{e3}{e-mail: weishw@lzu.edu.cn}
\thankstext{e4}{e-mail: liuyx@lzu.edu.cn, corresponding author}

\institute{
School of Physical Science and Technology, Southwest University, Chongqing 400715, China \label{addr1}
\and Department of Physics, Nanchang University, Nanchang 330031, China\label{addr2}
\and Center for Relativistic Astrophysics and High Energy Physics, Nanchang University, Nanchang 330031, China\label{addr3}
\and Institute of Theoretical Physics $\&$ Research Center of Gravitation, Lanzhou University, Lanzhou 730000, China\label{addr4}
\and Joint Research Center for Physics, Lanzhou University and Qinghai Normal University, Lanzhou 730000 and Xining 810000, China\label{addr5}
 }

\maketitle

\begin{abstract}

A novel four-dimensional Einstein-Gauss-Bonnet gravity was formulated by D. Glavan and C. Lin [Phys. Rev. Lett. 124, 081301 (2020)], which is intended to bypass the Lovelock's theorem and to yield a non-trivial contribution to the four-dimensional gravitational dynamics. However, the validity and consistency of this theory has been called into question recently.  We study a static and spherically symmetric black hole charged by a Born-Infeld electric field in the novel four-dimensional Einstein-Gauss-Bonnet gravity. It is found that the black hole solution still suffers the singularity problem, since particles incident from infinity can reach the singularity. It is also demonstrated that the Born-Infeld charged black hole may be superior to the Maxwell charged black hole to be a charged extension of the Schwarzschild-AdS-like black hole in this new gravitational theory. Some basic thermodynamics of the black hole solution is also analyzed. Besides, we regain the black hole solution in the regularized four-dimensional Einstein-Gauss-Bonnet gravity proposed by H. L\"u  and Y. Pang [arXiv:2003.11552].

\end{abstract}

\maketitle

\section{Introduction}

As the cornerstone of modern cosmology, Einstein's general relativity (GR) provides precise descriptions to a variety of phenomena in our universe. According to the powerful Lovelock's theorem \cite{Lovelock1971,Lovelock1972}, the only field equations of a four-dimensional (4D) metric theories of gravity that are second order or less are Einstein's equations with a cosmological constant. So in order to go beyond Einstein's theory, one usually modifies GR by adding some higher derivative terms of the metric or new degrees of freedom into the field equations, by considering other fields rather than the metric, or by extending to higher dimensions, etc. \cite{Clifton2012,Ishak2019}. A natural generalization of GR is the Lovelock gravity, which is the unique higher curvature gravitational theory that yields conserved second-order filed equations in arbitrary dimensions \cite{Lovelock1971,Lovelock1972}. Besides the Einstein-Hilbert term plus a cosmological constant, there is a Gauss-Bonnet (GB) term $ \mathcal{G}$ allowed in Lovelock's action in higher-dimensional spacetime. The GB term $ \mathcal{G}$ is quadratic in curvature and contributes ultraviolet corrections to Einstein's theory. Moreover, the GB term also appears in the low-energy effective action of string theory \cite{Zwiebach1985}. However, it is well known that the GB term is a total derivative in four dimensions, so it does not contribute to the gravitational dynamics. In order to generate a nontrivial contribution, one usually couples the GB term to a scalar field \cite{Odintsov2020,Odintsov2020a}.

In recent Refs.~\cite{Tomozawa2011,Cognola2013,Glavan2020}, the authors suggested that by rescaling the GB coupling constant $\alpha \to \alpha/(D-4)$ with $D$ the number of spacetime dimensions, the theory can bypass the Lovelock's theorem and the GB term can yield a non-trivial contribution to the gravitational dynamics in the limit $D\to 4$. This theory, now dubbed as the novel 4D Einstein-Gauss-Bonnet (EGB) gravity, will give rise to corrections to the dispersion relation of cosmological tensor and scalar modes, and is practically free from singularity problem in Schwarzschild-like black holes~\cite{Glavan2020}. The novel 4D EGB gravity has drawn intensive attentions recently \cite{Guo2020,Hegde2020,Zhang2020b,Singh2020,Konoplya2020a,Zhang2020a,Roy2020,HosseiniMansoori2020,Wei2020a,Singh2020a,Churilova2020,Islam2020,Mishra2020,Liu2020,Konoplya2020,HeydariFard2020,EslamPanah2020,NaveenaKumara2020,Aragon2020,Cuyubamba2020,Zhang2020,Yang2020,Ying2020,Doneva2020,Malafarina2020,Li2020,Kobayashi2020,Casalino2020,Rayimbaev2020,Liu2020a,Liu2020b,Bonifacio2020,Zeng2020,Ge2020}.  

However, there are also some debates on whether the novel gravity is a consistent and well-defined theory in four dimensions \cite{Ai2020,Gurses2020,Mahapatra2020,Shu2020,Tian2020,Arrechea2020,Lu2020,Fernandes2020a,Hennigar2020}. Such as, Gurses et al. pointed out that the novel 4D EGB gravity does not admit a description in terms of a covariantly-conserved rank-2 tensor in four dimensions and the dimensional regularization procedure is ill-defined, since one part of the GB tensor, the Lanczos-Bach tensor, always remains higher dimensional \cite{Gurses2020}. So generally speaking, the theory only makes sense on some selected highly-symmetric spacetimes, such as the FLRW spacetime and static spherically symmetric spacetime. Mahapatra found that in the limiting $D \to 4$ procedure, the theory gives unphysical divergences in the on-shell action and surface terms in four dimensions \cite{Mahapatra2020}. Shu showed that at least the vacuum of the theory is either unphysical or unstable, or has no well-defined limit as $D \to 4$ \cite{Shu2020}. Based on studying the evolution of the homogeneous but anisotropic universe described by the Bianchi type I metric, Tian and Zhu found that the novel 4D EGB gravity with the dimensional-regularization approach is not a complete theory \cite{Tian2020}. On the other hand, some regularized versions of 4D EGB gravity were also proposed by \cite{Lu2020,Fernandes2020a,Hennigar2020}, and the resulting theories belong to the family of Horndeski gravity. 

New black hole solutions in the novel 4D EGB gravity were also reported recently. In Ref.~\cite{Fernandes2020}, Fernandes generalized the Reissner-Nordstr\"om (RN) black hole with Maxwell electric field coupling to the gravity. The authors in Refs.~\cite{Kumar2020,Kumar2020a} considered the Bardeen-like and Hayward-like black holes by interacting the new gravity with nonlinear electrodynamics. The rotating black holes were also investigated in this new theory \cite{Wei2020,Kumar2020b}. Some other black hole solutions were discussed in the Refs.~\cite{Ghosh2020,Ghosh2020a,Ghosh2020b}.

A well-known nonlinear electromagnetic theory, called Born-Infeld (BI) electrodynamics, was proposed by Born and Infeld in 1934  \cite{Born1934}. They introduced a force by limiting the electromagnetic field strength analogy to a relativistic limit on velocity, and regularized the ultraviolet divergent self-energy of a point-like charge in classical dynamics. Since it was found that the BI action can be generated in some limits of string theory \cite{Fradkin1985}, the BI electrodynamics has become more popular. As an extension of RN black holes in Einstein-Maxwell theory, charged black hole solutions in Einstein-Born-Infeld (EBI) theory has received some attentions in recent years, see Refs. \cite{Garcia1984,Demianski1986,Dey2004,Cai2004a,Aiello2004,Miskovic2008,Gunasekaran2012,Dehyadegari2018} for examples.

In this work, we are interested in generalizing the static spherically symmetric black hole solutions charged by the BI electric field in the novel 4D EGB gravity. The paper is organized as follows. In Sec.~\ref{solution}, we solve the theory to get an exact black hole solution. In Sec.~\ref{thermodynamics}, we study some basic thermodynamics of the black hole and analyze its local and global stabilities. In Sec.~\ref{Regain_solution}, we regain the black hole  solution in the regularized 4D EGB gravity based on the work of H. L\"u  and Y. Pang \cite{Lu2020}. Finally, brief conclusions are presented.

 \section{BI Black Hole solution in the novel 4D EGB gravity}\label{solution}

We start from the action of the $D$-dimensional EGB gravity minimally coupled to the BI electrodynamics in the presence of a negative cosmological constant $\Lambda=-\frac{(D-1)(D-2)}{2 l^2}$,
\beqn
\mathcal{S}&=&\frac{1}{16\pi}\int{}d^Dx\sqrt{-g}\lt(R-2\Lambda+\frac{\alpha}{D-4} \mathcal{G}+\mathcal{L}_{BI} \rt),
\label{action}
\eeqn
where the GB term is $\mathcal{G}=R^2-4 R_{\mu \nu } R^{\mu \nu }+R_{\mu \nu \rho \sigma } R^{\mu \nu \rho \sigma }$, and the Lagrangian of the BI electrodynamics reads
\beq
\mathcal{L}_{BI}=4\beta^2\left(1-\sqrt{1+\frac{F_{\mu\nu}F^{\mu\nu}}{2\beta^2}}\right).
\label{BI_action}
\eeq
Here $\beta>0$ is the BI parameter and it is the maximum of the electromagnetic field strength. The Maxwell electrodynamics is recovered in the limit $\beta\to\infty$.

The equations of motion of the theory can be obtained by varying the action with respect to the metric field $g_{\mu\nu}$ and the gauge field $A_{\mu}$:
\beqn
G_{\mu\nu}+\Lambda g_{\mu\nu}+\frac{\alpha}{D-4}\Big(\frac{\delta  \mathcal{G}}{\delta g^{\mu\nu}}-\frac{1}{2}g_{\mu\nu} \mathcal{G}\Big)&&\nn\\
+\lt(\frac{\delta \mathcal{L}_{BI}}{\delta g^{\mu\nu}}-\frac{1}{2}g_{\mu\nu}\mathcal{L}_{BI}\rt)&=&0,\label{EoM1}\\
\pt_\mu\Bigg(\frac{\sqrt{-g}F^{\mu\nu}}{\sqrt{1+\frac{F_{\rho\sigma}F^{\rho\sigma}}{2\beta^2}}}\Bigg)&=&0,\label{EoM2}
\eeqn
where
\beqn
\frac{\delta  \mathcal{G}}{\delta g^{\mu\nu}}&=&2RR_{\mu\nu}+2R_\mu{}^{\rho\sigma\lambda}R_{\nu\rho\sigma\lambda}-4R_{\mu\lambda}R^\lambda{}_\nu\nn\\
&&-4R^{\rho\sigma}R_{\mu\rho\nu\sigma},\\
\frac{\delta \mathcal{L}_{BI}}{\delta g^{\mu\nu}}
      &=&\frac{2F_{\mu\lambda} {F^{\lambda{}}}_{\nu}}
              {\sqrt{1+\frac{F_{\rho\sigma}F^{\rho\sigma}}{2\beta^2}}}.
\eeqn

Here we consider a static spherically symmetric metric ansatz in $D$-dimensional spacetime
\beq
ds^2=-a(r)e^{-2b(r)} dt^2+\frac{dr^2}{a(r)}+r^2d\Omega^2_{D-2},
\label{metric}
\eeq
where $d\Omega^2_{D-2}$ represents the metric of a $(D-2)$-dimensional unit sphere.  Correspondingly, the vector potential is assumed to be $A=\Phi(r)dt$.

Instead of solving the equations of motion directly, it would be more convenient to start from the following reduced action to get the solution,  i.e.,
\beqn
\mathcal{S}&=&\frac{\Sigma_{D-2}}{16\pi}\int{}dtdr (D-2) e^{-b}\Bigg[\bigg(r^{D-1}\psi(1+\alpha(D-3)\psi)\nn\\
&&+\frac{r^{D-1}}{l^2}\bigg)'+\frac{4\beta^2 r^{D-2}}{D-2}\bigg(1-\sqrt{1-{\beta^{-2} }e^{2b}\Phi'^2} \bigg) \Bigg],
\eeqn
where the prime denotes the derivative respect to the radial coordinate $r$, $\Sigma_{D-2}={2\pi^{\fc{D-1}{2}}}/{\Gamma[\frac{D-1}{2}]}$ is the area of a unit $(D-2)$-sphere, and $\psi(r)=(1-a(r))/r^2$. By varying the action with respect to $a(r)$, $b(r)$ and $\Phi(r)$, one can easily obtain the field equations and hence the solutions in arbitrary dimensions. Here we are interested in the case of $D=4$. The solution can be found in a closed form:
\beqn
b(r)&=&0,\label{solution_b}\\
\Phi(r)&=&\fc{Q}{r}{~}_2F_1\lt(\fc{1}{4},\fc{1}{2},\fc{5}{4},-\fc{Q^2}{\beta^2 r^4}\rt),\label{solution_phi}\\
a(r)&=&1+\fc{r^2}{2\alpha}\Bigg\{1\pm\Bigg[1+4\alpha\Bigg(\fc{2M}{r^3}-\fc{1}{l^2}\nn\\
&&-\fc{2\beta^2}{3}\bigg(1-\sqrt{1+\fc{Q^2}{\beta^2 r^4}}\bigg)- \fc{4Q}{3r^3}\Phi(r)\Bigg) \Bigg]^{\fc{1}{2}}\Bigg\},\label{sol_metric_function}
\eeqn
where we have chosen the integration constants to recover a proper asymptotic limit, and ${}_2F_1$ is the hypergeometric function. 

In the limit of $\beta\to\infty$, the solution recovers the Maxwell charged RN-AdS-like black hole solution in the novel 4D EGB gravity \cite{Fernandes2020},
\beq
a(r)=1+\fc{r^2}{2\alpha}\lt[1\pm\sqrt{1+4\alpha\lt(\fc{2M}{r^3}-\fc{Q^2}{r^4}-\fc{1}{l^2}\rt) }\rt].\label{sol_metric_function_RNEGB}
\eeq
Moreover, the Schwarzschild-AdS-like black hole solution in the novel 4D EGB gravity is recovered by closing the electric charge \cite{Glavan2020},
\beq
a(r)=1+\fc{r^2}{2\alpha}\lt[1\pm\sqrt{1+4\alpha\lt(\fc{2M}{r^3}-\fc{1}{l^2}\rt) }\rt].\label{sol_metric_function_ScEGB}
\eeq

On the other hand, in the large distance, the metric function $a(r)$ (\ref{sol_metric_function}) has the asymptotic behavior
\beqn
a(r)&=& 1+\frac{r^2}{2\alpha}\lt(1\pm\sqrt{1-\fc{4\alpha}{l^2}}\rt)\pm\fc{2M r-Q^2}{r^2\sqrt{1-\fc{4\alpha}{l^2}}}\nn\\
&&+\mathcal{O}({r^{-4}}).
\eeqn
So in order to have a real metric function in large distance, it is clear that we require $0<\alpha\leq l^2/4$ or $\alpha<0$. Further, by taking the small $\alpha$ limit, the solution of the plus-sign branch reduces to a RN-AdS solution with a negative gravitational mass and imaginary charge, and only the minus-sign branch can recover a proper RN-AdS limit,
\beq
a(r)= 1-\fc{2M}{r}+\fc{Q^2}{r^2}+\fc{r^2}{l^2}+\mathcal{O}(\alpha).
\eeq
So we only focus on the solution of the minus-sign branch in the rest of the paper. 

\begin{figure}[htbp]
\centering
\subfigure[$~Q=0.8$]{\label{Pic_ar1}
\includegraphics[width=4cm,height=3.5cm]{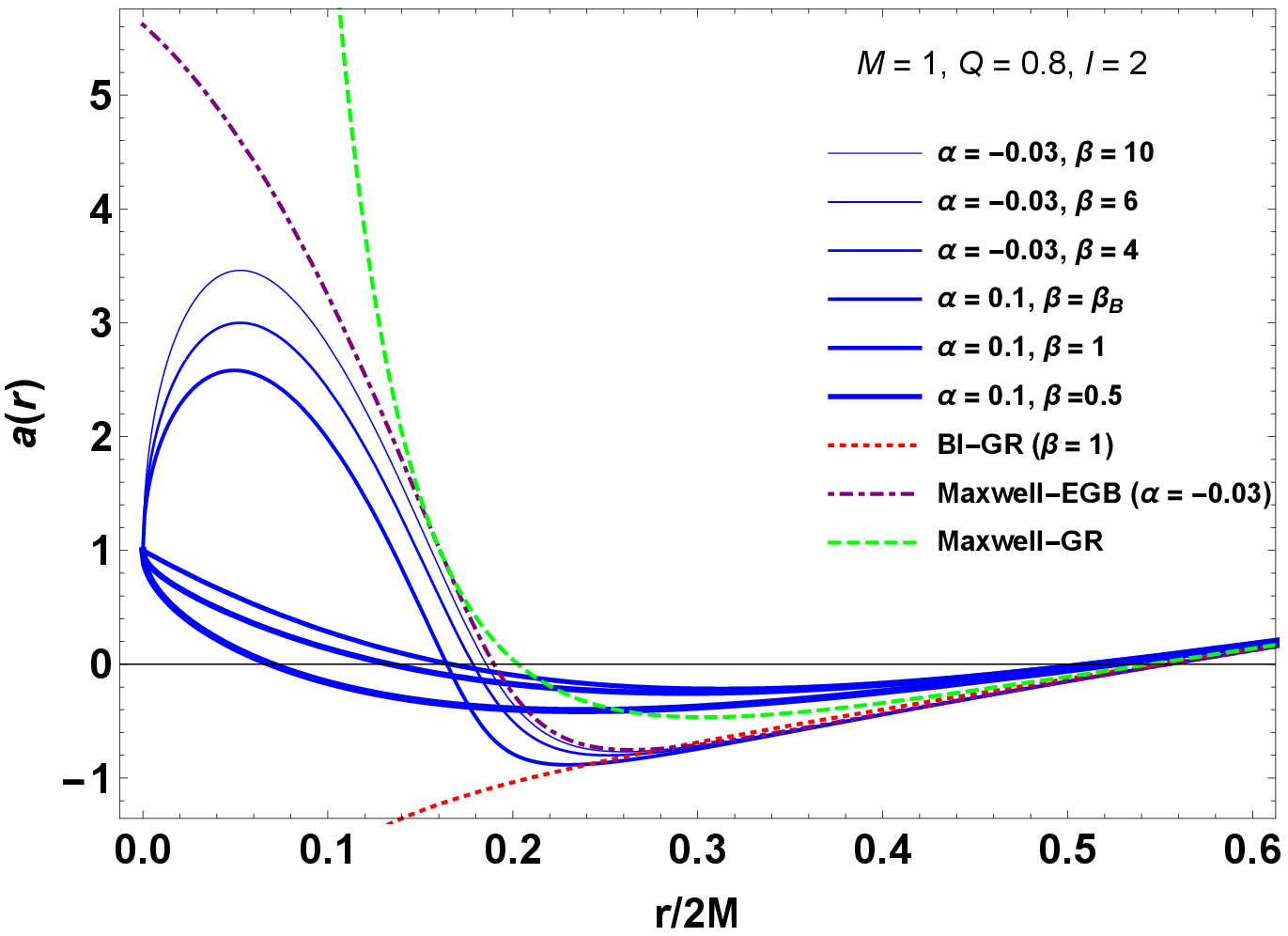}}
\subfigure[$~Q=1$]{  \label{Pic_ar_Q_1}
\includegraphics[width=4cm,height=3.5cm]{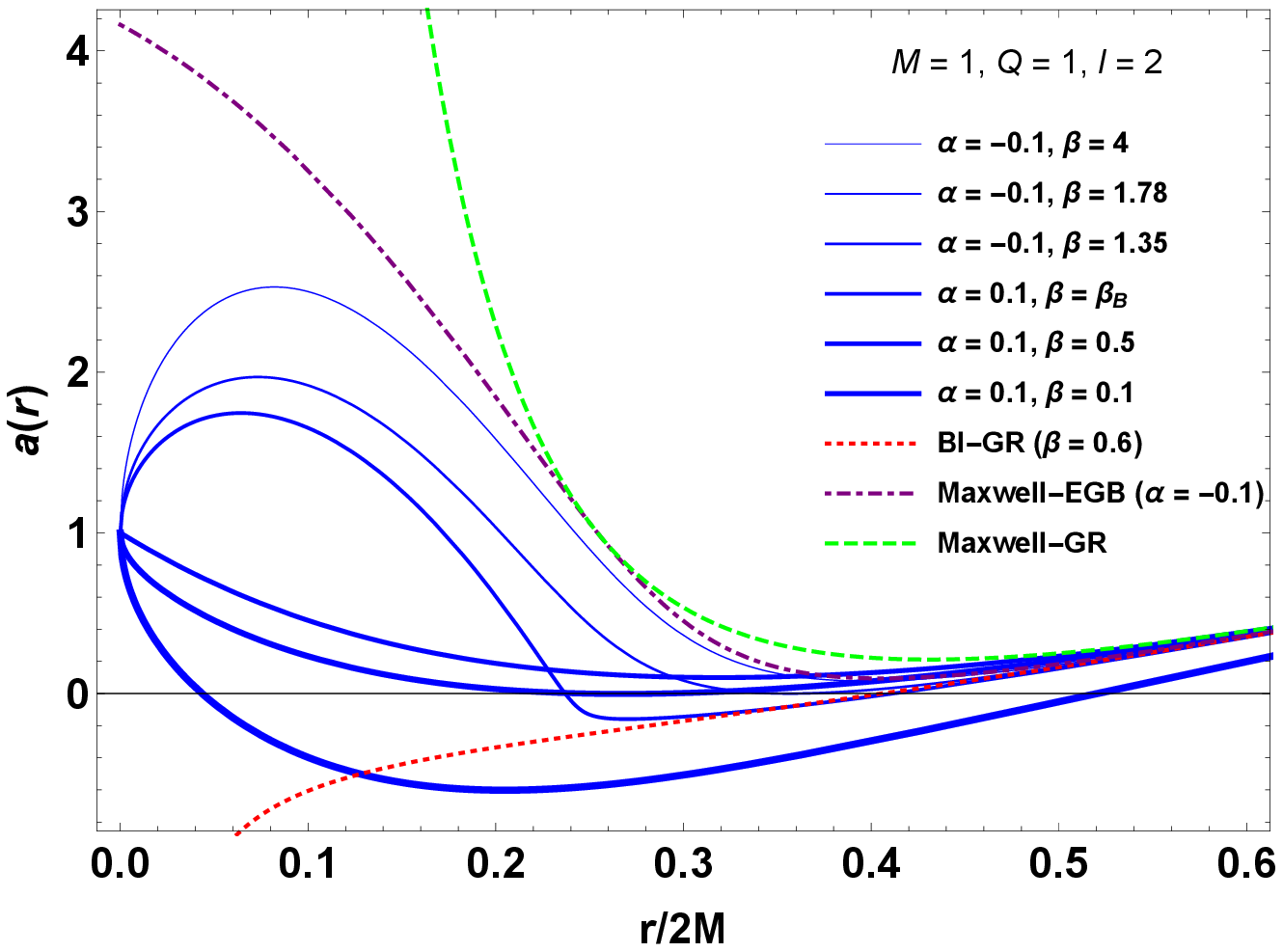}}\\
\caption{The metric function $a(r)$ for some parameter choices.
  }
\label{metric_function}
\end{figure}

We illustrate the minus-sign branch of the metric function $a(r)$ in Fig.~\ref{metric_function} for some parameter choices, where the solutions of the BI charged black hole in GR and the Maxwell charged black holes in the novel 4D EGB gravity and GR are also plotted for comparison. As shown in the figures, similar to the RN black hole, the solution (\ref{sol_metric_function}) can have zero, one, or two horizons depending on the parameters. However, the metric function $a(r)$ always approaches a finite value $a(0)=1$ at the origin $r\to0$ in the novel 4D EGB gravity.  This property is different from the BI charged AdS black hole in EBI theory~\cite{Gunasekaran2012}, where there are different types of solutions relying on the parameters: a ``Schwarzschild-like" type for $\beta<\beta_B$ (only one horizon with $a(0)\to-\infty$), an ``RN" type for $\beta>\beta_B$ (naked singularity, one or two horizons with $a(0)\to+\infty$),  and a ``marginal" type for $\beta=\beta_B$ (naked singularity or one horizon with a finite value $a(0)=1-2\beta_B Q$). Interestingly, as shown in Fig. \ref{Pic_ar_Q_1}, even the solution coupled to the Maxwell field becomes naked singularities, the ones coupled to the BI field are still regular black holes.

In the small distance limit,  the hypergeometric function behaves like
\beqn
{~}_2F_1\lt(\fc{1}{4},\fc{1}{2},\fc{5}{4},-\fc{Q^2}{\beta^2 r^4}\rt)&=&\sqrt{\fc{\beta}{\pi Q}}\fc{\Gamma^2(1/4) }{4}r-\fc{\beta }{Q}r^2\nn\\
&&+\mathcal{O}(r^6).
\eeqn
Then the formula in the square root of $a(r)$ can be rearranged as
\beqn
1&+&4\alpha\Bigg[\fc{2M}{r^3}\lt(1-\sqrt{\fc{\beta}{\beta_B}} \rt)  +\fc{4\beta Q}{3 r^2}-\fc{1}{l^2}\nn\\
&&-\fc{2\beta^2}{3}\lt(1-\sqrt{1+\fc{Q^2}{\beta^2 r^4}}\rt)\Bigg]+\mathcal{O}(r^2),
\eeqn
where $\beta_B\equiv\frac{36 \pi M^2}{\Gamma(1/4)^4 Q^3}$. One special case appears at $\beta=\beta_B$, where the metric function $a(r)$ (\ref{sol_metric_function}) in the small distance limit behaves like
\beq
a(r)=1-\sqrt{\fc{2\beta_B Q}{\alpha}}r+\fc{r^2}{2\alpha}+\mathcal{O}(r^3),
\label{MetFunc_small_r_limit1}
\eeq
and the Ricci scalar $R\approx\sqrt{\fc{2\beta_B Q}{\alpha}}\fc{6}{r}$ is divergent. However, if $\beta\neq\beta_B$, $a(r)$  behaves like
\beq
a(r)= 1-\fc{\sqrt{K} r^\fc{1}{2}}{\alpha}-\fc{\beta Q r^\fc{3}{2}}{\sqrt{K}}+\fc{r^2}{2\alpha}+\mathcal{O}(r^{\fc{5}{2}}),
\label{MetFunc_small_r_limit2}
\eeq
where $ K=2 M \alpha \lt(1-\sqrt{\fc{\beta}{\beta_B}}\rt)$. Now the Ricci scalar reads $R\approx\fc{15}{8\alpha}\fc{\sqrt{K}}{ r^{3/2}}$. So, by considering Eqs. (\ref{MetFunc_small_r_limit1}) and (\ref{MetFunc_small_r_limit2}), we require $\alpha>0$, $0<\beta\leq\beta_B$ or  $\alpha<0$, $\beta>\beta_B$ in order to have a real metric function.

\begin{figure}[htbp]
\centering
\subfigure[$~\alpha<0, Q=0.8$]{\label{parameter_constraints_negative}
\includegraphics[width=4cm,height=3.5cm]{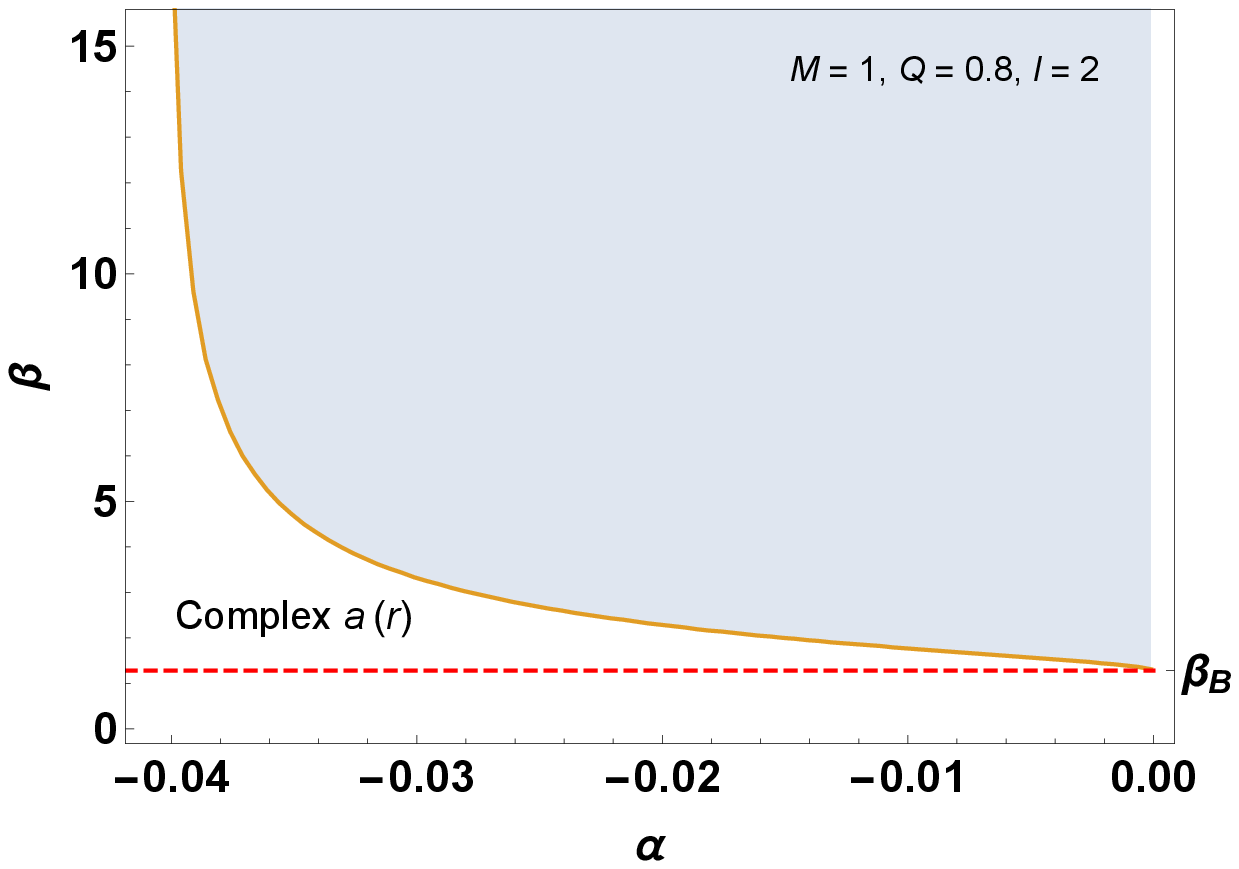}}
\subfigure[$~\alpha>0, Q=0.8$]{  \label{parameter_constraints_positive}
\includegraphics[width=4cm,height=3.5cm]{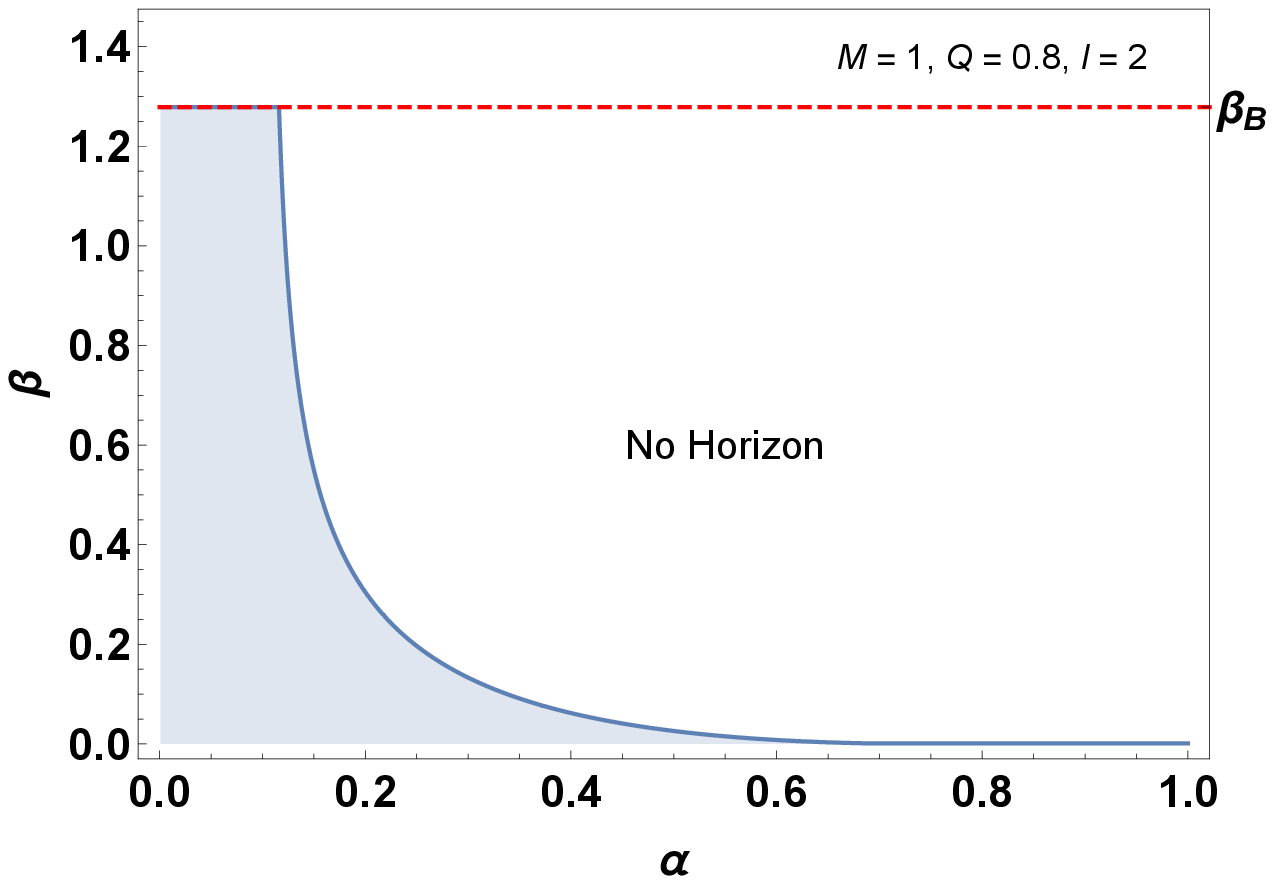}}
\subfigure[$~\alpha<0, Q=1$]{\label{parameter_constraints_negative2}
\includegraphics[width=4cm,height=3.5cm]{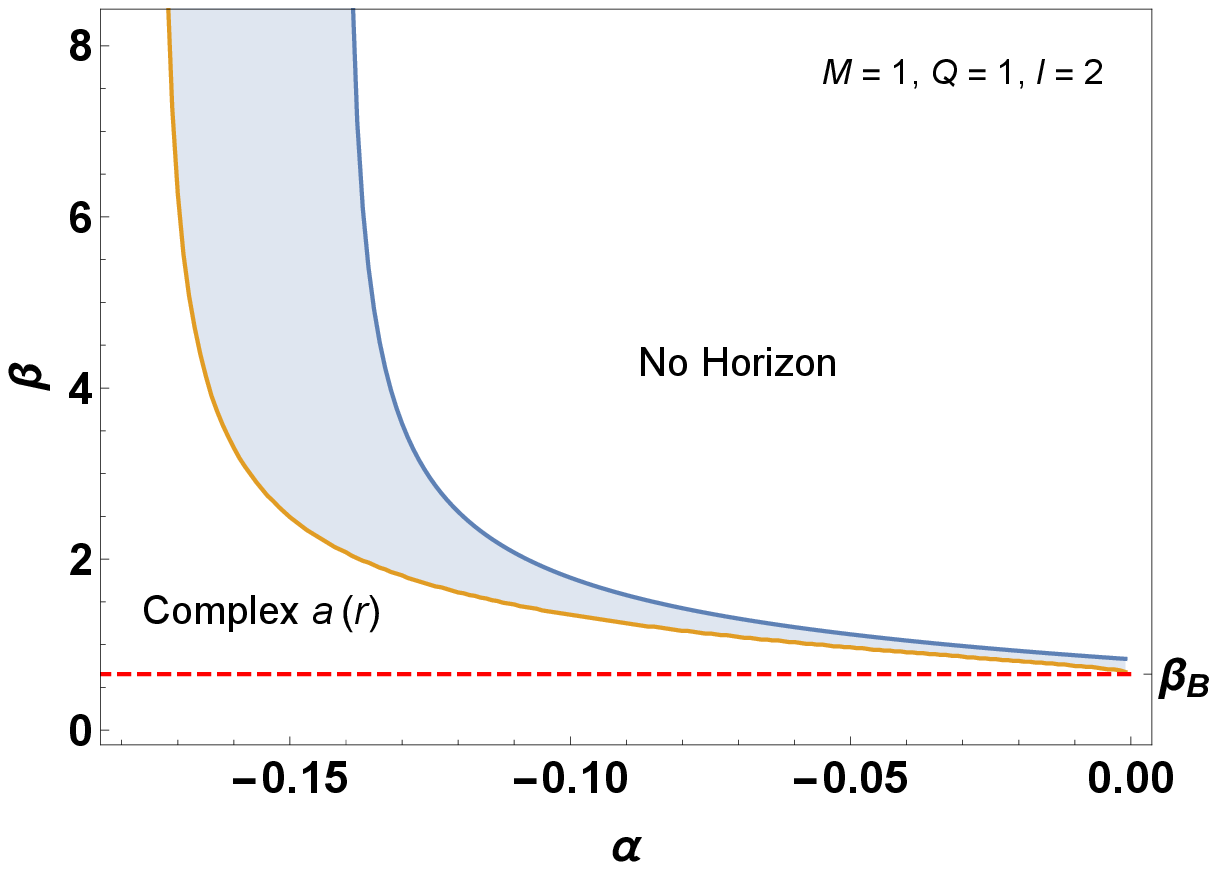}}
\subfigure[$~\alpha>0, Q=1$]{  \label{parameter_constraints_positive2}
\includegraphics[width=4cm,height=3.5cm]{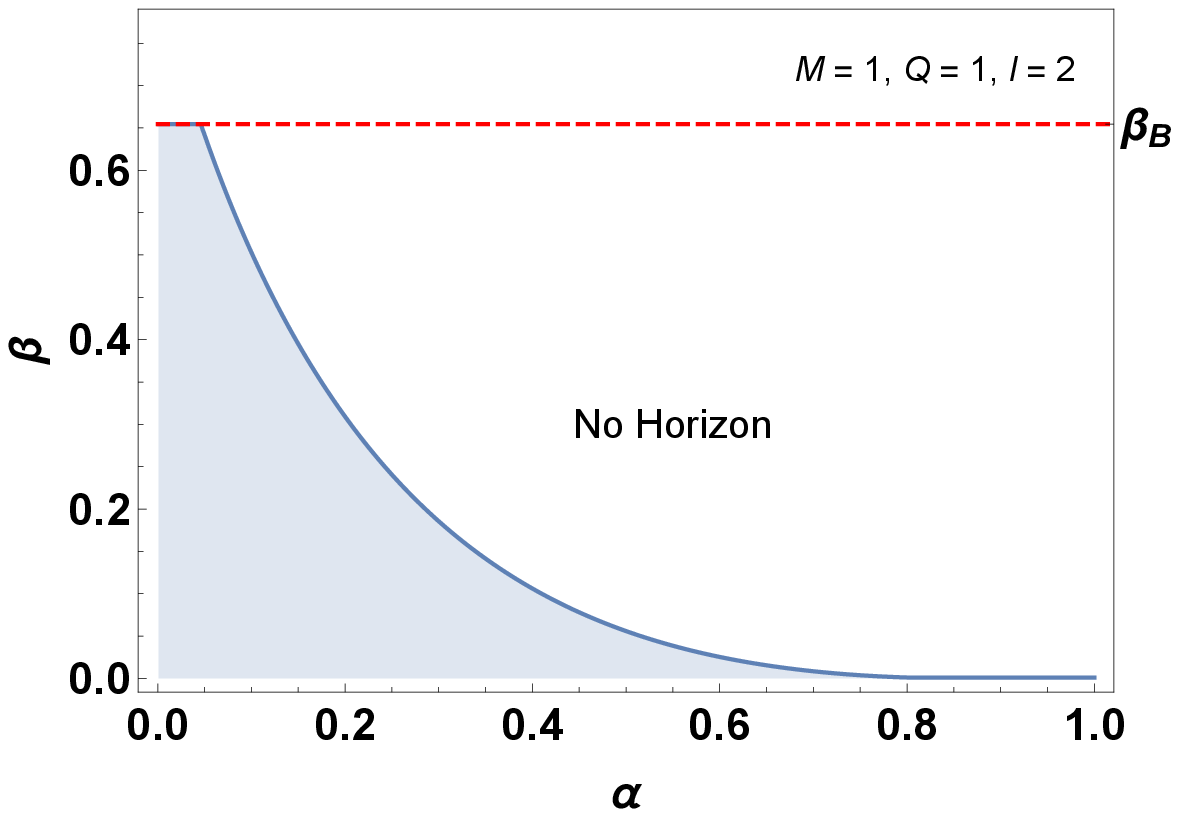}}
\caption{The parameter space for the case of $M=1$, $Q=0.8,~1$, and $l=2$, where the shaded area represents available parameter region.}
\label{parameter_constraints}
\end{figure}

The above constraints $0<\alpha\leq l^2/4, 0<\beta\leq\beta_B$ or $\alpha<0, \beta>\beta_B$ are not enough to guarantee a real metric function in the whole parameter space. More accurate constraints can be worked out by requiring the formula under the square root of the metric function $a(r)$ (\ref{sol_metric_function}) to be positive. Moreover, in order to ensure the existence of the black hole horizon, there are further constraints on $\alpha$ and $\beta$, as shown in Fig.~\ref{metric_function}.  Due to the complicated form of $a(r)$,  we fail to get the analytic full constraints. Nevertheless, we can numerically solve the constraints for fixed $Q$, $M$, and $l$. For example, we show the results for the cases of $Q=0.8$ and $Q=1$ with $M=1$ and $l=2$ in Fig.~\ref{parameter_constraints}. As shown in the figures, regular black hole solution only exists in the shaded parameter region, and the solution turns into a naked singularity above the shaded region. Especially, on the border, the two horizons degenerate and the solution corresponds to an extremal black hole.

Since the leading term ${8\alpha M}/{r^3}$ under the square root of Eq.~(\ref{sol_metric_function_ScEGB}) is negative at short radius for $\alpha<0$, a real metric function is present only for positive $\alpha$ \cite{Glavan2020}. However, for the charged solution (\ref{sol_metric_function_RNEGB}), the leading term $-{4\alpha Q^2}/{r^4}$ under the square root is always negative at small radius for positive $\alpha$, so one requires $\alpha<0$ to get a real metric function. Therefore, by ``throwing" electric charges into the Schwarzschild-AdS-like black hole (\ref{sol_metric_function_ScEGB}) with positive $\alpha$, it would not be deformed into the Maxwell charged RN-AdS-like black hole (\ref{sol_metric_function_RNEGB}) with positive $\alpha$. However, as illustrated in Fig.~\ref{parameter_constraints}, the BI charged black hole solution (\ref{sol_metric_function}) can be real for both positive and negative $\alpha$. In the limit of $Q\to 0$, $\beta_B=\frac{36 \pi M^2}{\Gamma(1/4)^4 Q^3}\to \infty$, the BI charged black hole with positive $\alpha$ can be continuously deformed into the Schwarzschild-AdS-like black hole (\ref{sol_metric_function_ScEGB}). This is a hint that the BI charged black hole may be superior to the Maxwell charged black hole in the novel 4D EGB gravity.

Note that the metric (\ref{MetFunc_small_r_limit2}) approaches a finite value $a(0)=1$ as $r\to0$, and it is similar to the behavior of (\ref{sol_metric_function_ScEGB}) in Ref.~\cite{Glavan2020} for positive $\alpha$. So in this case, an infalling particle would feel a repulsive gravitational force when it approaches the singular point $r=0$. However, for the solution with negative $\alpha$, the infalling particle would feel an attractive gravitational force in short distance. Interestingly, for the critical case (\ref{MetFunc_small_r_limit1}) with $\alpha>0$ and $\beta=\beta_B$, there is a maximal repulsive gravitational force when the particle approaches the singular point. However, whether the particle can reach the singularity or not depends on its initial conditions. 

\begin{figure}[htbp]
\centering
\subfigure[$~\alpha=0.1, \beta=1$]{
\includegraphics[width=4cm,height=3.5cm]{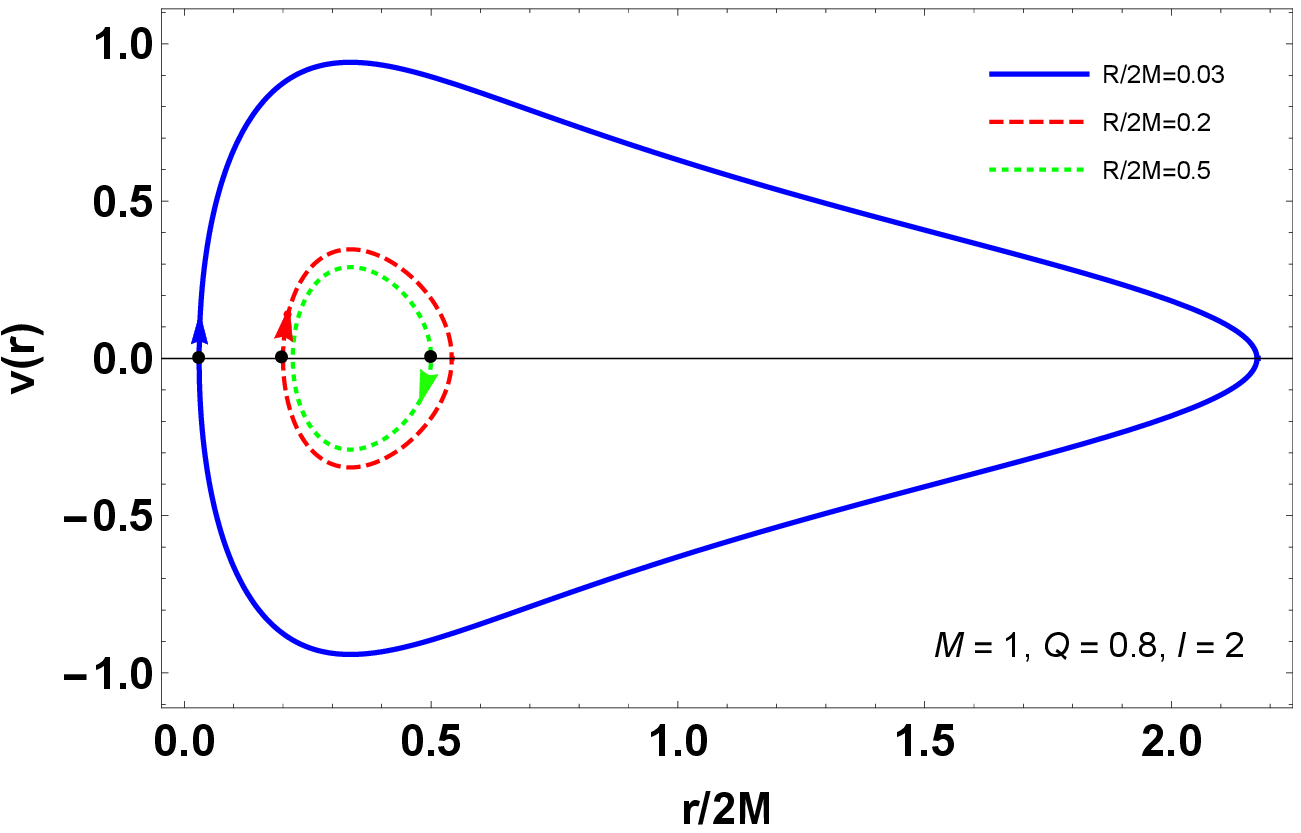}}
\subfigure[$~\alpha=-0.03, \beta=4$]{ 
\includegraphics[width=4cm,height=3.5cm]{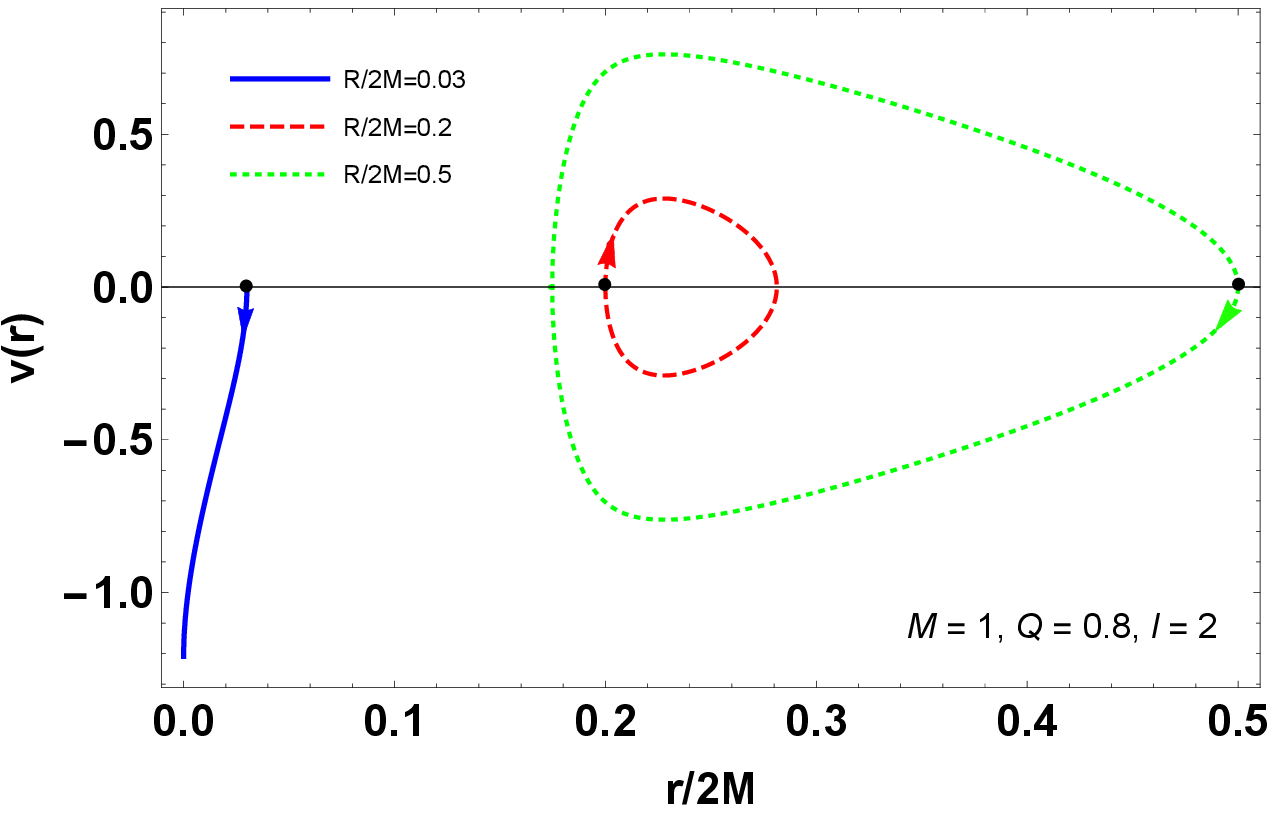}}\\
\caption{The velocity $v(r)$ of an infalling particle for starting at rest at different radii $R$, where the dot represents the initial position and the arrow represents the direction in which the velocity starts to increase.}
\label{velocity}
\end{figure}

When the particle starts at rest at the radius $R$ and freely falls radially toward the black hole, its velocity is given by  \cite{Misner1973}
\beq
\fc{dr}{d\tau}=\pm \sqrt{a(R)-a(r)},
\eeq
where $a(R)=\tilde{E}^2$ with $\tilde{E}$ the energy per unit rest mass, and the plus (minus) sign refers to the infalling (outgoing) particle. For simplicity we consider the case of asymptotically Minkowski space, i.e., $l\to\infty$. As illustrated in Fig.~\ref{velocity}, if the the radius $R$ is finite, the particle can not reach the singularity for the solution with positive $\alpha$, but it can reach the singularity in short distance for the solution with negative $\alpha$, since the particle feels an attractive gravitational force when it gets close to the singularity. However, if the particle starts at rest at infinity, i.e., $\tilde{E}^2=a(R\to\infty)=1$, since $a(r\to0)=1$ as seen in Eqs. (\ref{MetFunc_small_r_limit1}) and (\ref{MetFunc_small_r_limit2}), it will just reach the singularity with zero speed \cite{Arrechea2020}. So if the particle has a kinetic energy at infinity, namely, $\tilde{E}^2=a(R\to\infty)>1$, it will reach the singularity with a nonzero speed. Therefore, in this sense, the black hole solution still suffers the singularity problem.

\section{Thermodynamics}\label{thermodynamics}

Black holes are widely believed to be thermodynamic objects, which possess the standard thermodynamic variables and satisfy the four laws of black hole thermodynamics. In this section, we explore some basic thermodynamics of the BI charged black hole in the novel 4D EGB gravity.

\begin{figure}[htbp]
\centering
\subfigure[$~\alpha>0$]{  \label{pic_mass_positive}
\includegraphics[width=4cm,height=3.5cm]{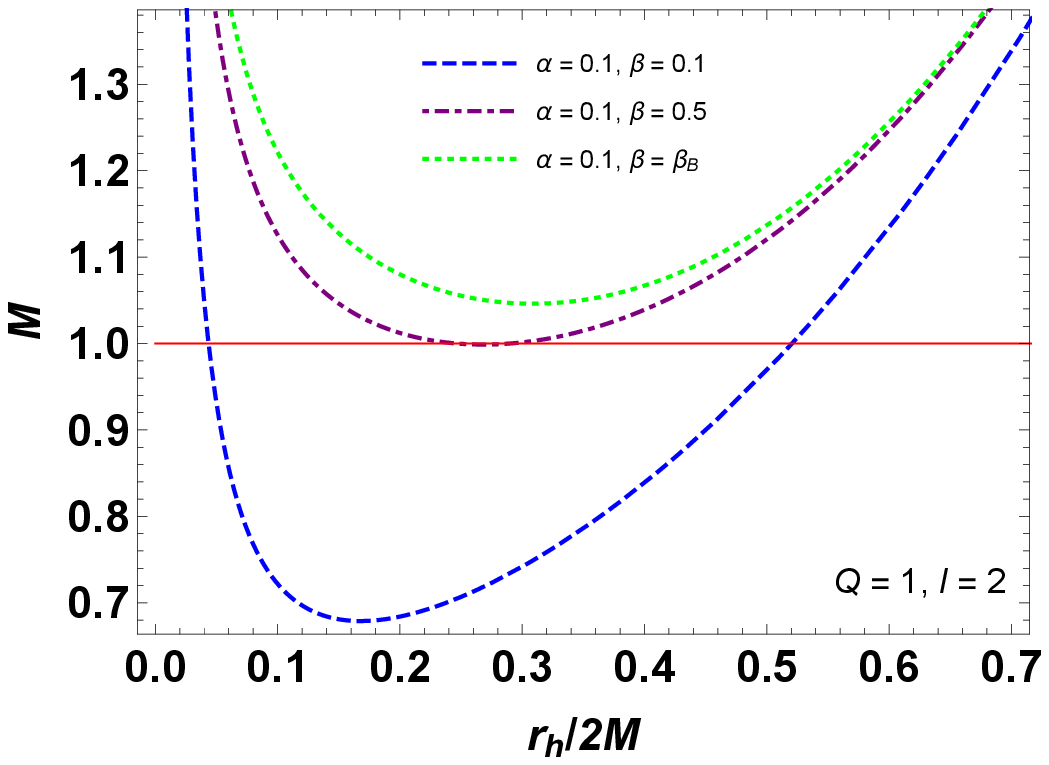}}\subfigure[$~\alpha<0$]{\label{pic_mass_negative}
\includegraphics[width=4cm,height=3.5cm]{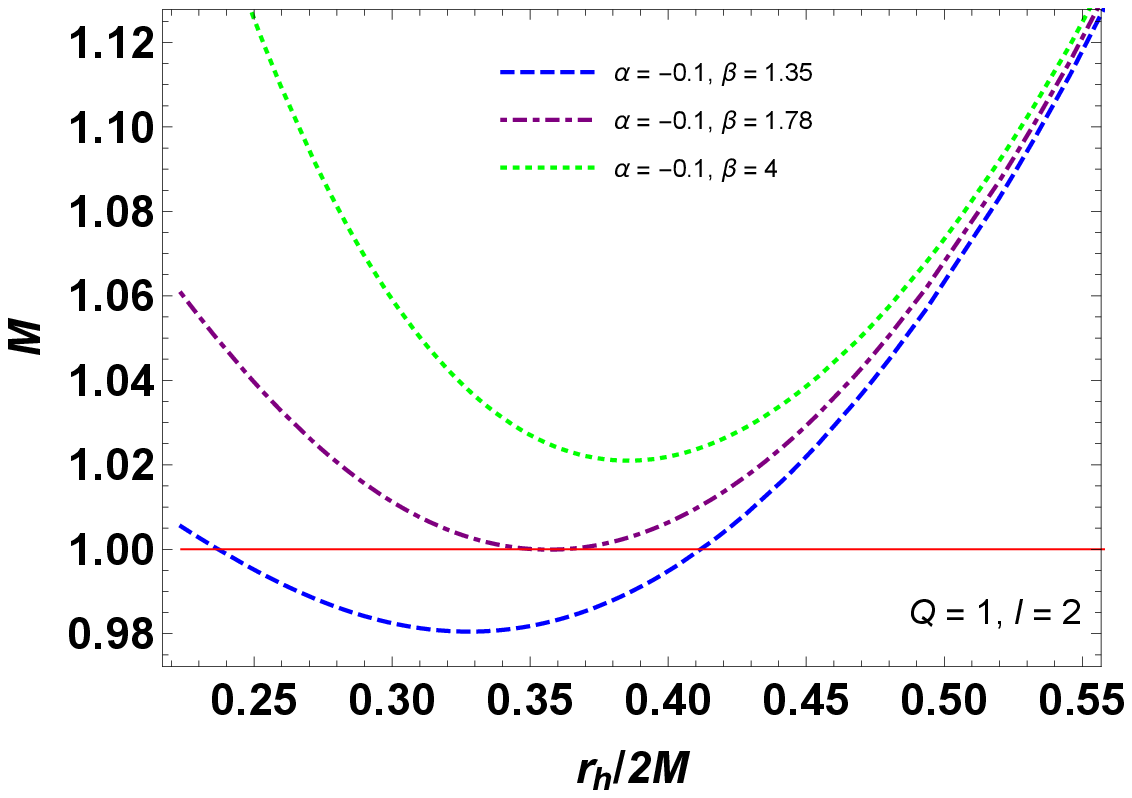}}\\
\caption{The mass of the black hole for positive $\alpha$ (a) and  negative $\alpha$  (b).   }
\label{pic_mass}
\end{figure}

By setting $a(r)=0$ in Eq.~(\ref{sol_metric_function}), the black hole mass can be expressed with the radius of the event horizon $r_h$,  namely,
\beqn
M&=&\fc{r_h}{2}\lt[1+\fc{r_{h}^2}{l^2}+\fc{\alpha}{r^2}+\fc{2\beta^2r_{h}^2}{3}\lt(1-\sqrt{1+\fc{Q^2}{\beta^2r_{h}^4}} \rt)\rt.\nn\\
&&\lt.\quad\quad+\fc{4Q^2}{3r^2}{~}_2F_1\lt(\fc{1}{4},\fc{1}{2},\fc{5}{4},-\fc{Q^2}{\beta^2 r_{h}^4}\rt) \rt],
\label{BH_mass}
\eeqn
where the horizon radius has to satisfy $r_h>\sqrt{-2\alpha}$ for negative $\alpha$, since the square root of metric function (\ref{sol_metric_function}) must be positive. So there is a minimum horizon radius $r_h^*=\sqrt{-2\alpha}$ for negative $\alpha$. As shown in Fig. \ref{pic_mass}, for a given mass, the number of horizons depends on the parameters $\alpha$ and $\beta$, and the corresponding parameter space is explicitly shown in Fig. \ref{parameter_constraints}. The critical case, i.e., an extremal black hole, just happens when $M'(r_h)=0$, which yields
\beq
1+\fc{3 r_h^2}{l^2}-\fc{\alpha }{r_h^2}+2 \beta ^2 r_h^2 \left(1-\sqrt{1+\fc{Q^2}{\beta ^2 r_h^4}}\right)=0.
\label{critical_condition}
\eeq

In the presence of the GB term and BI electrodynamics, the dimensionful GB coupling parameter $\alpha$ and BI parameter $\beta$ should be also treated as new thermodynamic variables \cite{Gunasekaran2012}. Now the generalized Smarr formula is given by
\beq
M=2(TS-VP+\mathcal{A} \alpha)+\Phi Q-\mathcal{B}\beta.
\label{smarr_formula}
\eeq
The electrostatic potential $\Phi$ given in (\ref{solution_phi}) is measured at the horizon $r_h$ in this formula.

The Hawking temperature can be obtained from the relation $T=\fc{\kappa}{2\pi}$ with the surface gravity  $\kappa=\lt.-\fc{1}{2}\fc{\pt g_{tt}}{\pt r}\rt|_{r=r_h}$, and reads
\beqn
T&=&\fc{1}{4\pi r_{h}}\Bigg[\fc{3r_{h}^4-3\alpha l^2+2\beta^2 l^2 r_{h}^4\lt(1-\sqrt{1+\fc{Q^2}{\beta^2 r_{h}^4}} \rt)}{l^2(r_{h}^2+2\alpha)}\nn\\
&&\quad\quad\quad+1  \Bigg].
\eeqn
The temperature vanishes for the extremal black hole, which can be easily proved with the critical condition (\ref{critical_condition}). In the large $\beta$ and small $\alpha$ limits, the temperature reduces to
\beqn
T&=&\fc{1}{4\pi r_{h}}\lt(1-\fc{Q^2}{r_{h}^2}+\fc{3r_{h}^2}{l^2} \rt)-\fc{\alpha}{4\pi r_+}\lt(\fc{6}{l^2}+\fc{3}{r_{h}^2}-\fc{2Q^2}{r_{h}^4} \rt)\nn\\
&&+\fc{Q^4}{16\pi \beta^2 r_{h}^7}+\mathcal{O}\left(\alpha^2,\fc{\alpha}{\beta^2},\fc{1}{\beta^4}\right),
\eeqn
where the first term is the standard Hawking temperature of the RN-AdS black hole, the second term is the leading order correction from the Gauss-Bonnet term, and the third term is the leading order Born-Infeld correction.

By following the approach of Ref. \cite{Cai1999}, the entropy of the black hole can be worked out from
\beq
S=\int{} \frac{1}{T}\lt(\fc{\pt M}{\pt r_h}  \rt)_{P,Q,\alpha,\beta} dr_{h} + S_0,
\eeq
where $S_0$ is an integration constant. Then it yields
\beq
S=\pi r_h^2+2\pi\alpha\ln r_h^2+S_0=\fc{A_h}{4}+2\pi\alpha \ln\fc{A_h}{A_0},
\label{BH_entropy}
\eeq
where $A_h=4\pi r_h^2$ is the horizon area and $A_0$ is a constant with dimension of area, which is not determined from first principle \cite{Lu2020}. By considering that the black entropy is generally independent of the black hole charge and cosmological constant but is relevant to the GB coupling parameter $\alpha$, which has the dimension of area,  we simply fix the undetermined constant as $A_0=4\pi |\alpha|$ \cite{Wei2020a}. Here a logarithmic term associated with the GB coupling parameter $\alpha$ arises as a subleading correction to the Bekenstein-Hawking area formula, which is universal  in some quantum theories of gravity \cite{Cai2010c}. Since the novel 4D EGB gravity is considered as a classical modified gravitational theory \cite{Glavan2020}, the logarithmic correction appears at classical level here.

The pressure $P=\fc{3}{8\pi l^2}$ is associated with the cosmological constant \cite{Kastor2009,Kubiznak2012}, and the corresponding thermodynamical volume $V$ is given by
\beq
V=\lt(\fc{\pt M}{\pt P}\rt)_{S,Q,\alpha,\beta}=\fc{4\pi r_h^3}{3}.
\eeq

By noting that the entropy is associated with the horizon $r_h$ and GB coupling parameter $\alpha$, the conjugate quantity $\mathcal{A}$ of $\alpha$ can be obtained from Eqs. (\ref{BH_mass}) and (\ref{BH_entropy}) as
\beqn
\mathcal{A}&=&\fc{1}{2r_h}+\lt[\fc{3r^3}{2l^2}+\fc{r^2-\alpha}{3r}+\beta^2r^3\lt(1-\sqrt{1+\fc{Q^2}{\beta^2 r^4}}\rt)  \rt]\nn\\
&&\times\fc{1-\ln\fc{r^2}{|\alpha|}}{r^2+2\alpha}.
\eeqn

Finally, we can calculate the BI vacuum polarization $\mathcal{B}$ from Eq. (\ref{BH_mass}):
\beq
\mathcal{B}\!=\!\fc{2r_h^3}{3}\lt(1\!-\!\sqrt{1+\fc{Q^2}{\beta^2 r^4}} \rt)+\fc{Q^2}{3\beta r}\!{~}_2F_1\!\lt(\fc{1}{4},\fc{1}{2},\fc{5}{4},-\fc{Q^2}{\beta^2 r_{h}^4}\rt).
\eeq

With all the above thermodynamic quantities at hand, it is easy to verify the validity of Smarr formula (\ref{smarr_formula}). Further, it is also straightforward to verify that the first law of black hole
\beq
dM=TdS-VdP+\mathcal{A}d \alpha+\Phi dQ-\mathcal{B}d\beta
\eeq
holds as expected.

\begin{figure}[htbp]
\centering
\subfigure[]{  \label{Sheat}
\includegraphics[width=4cm,height=3.5cm]{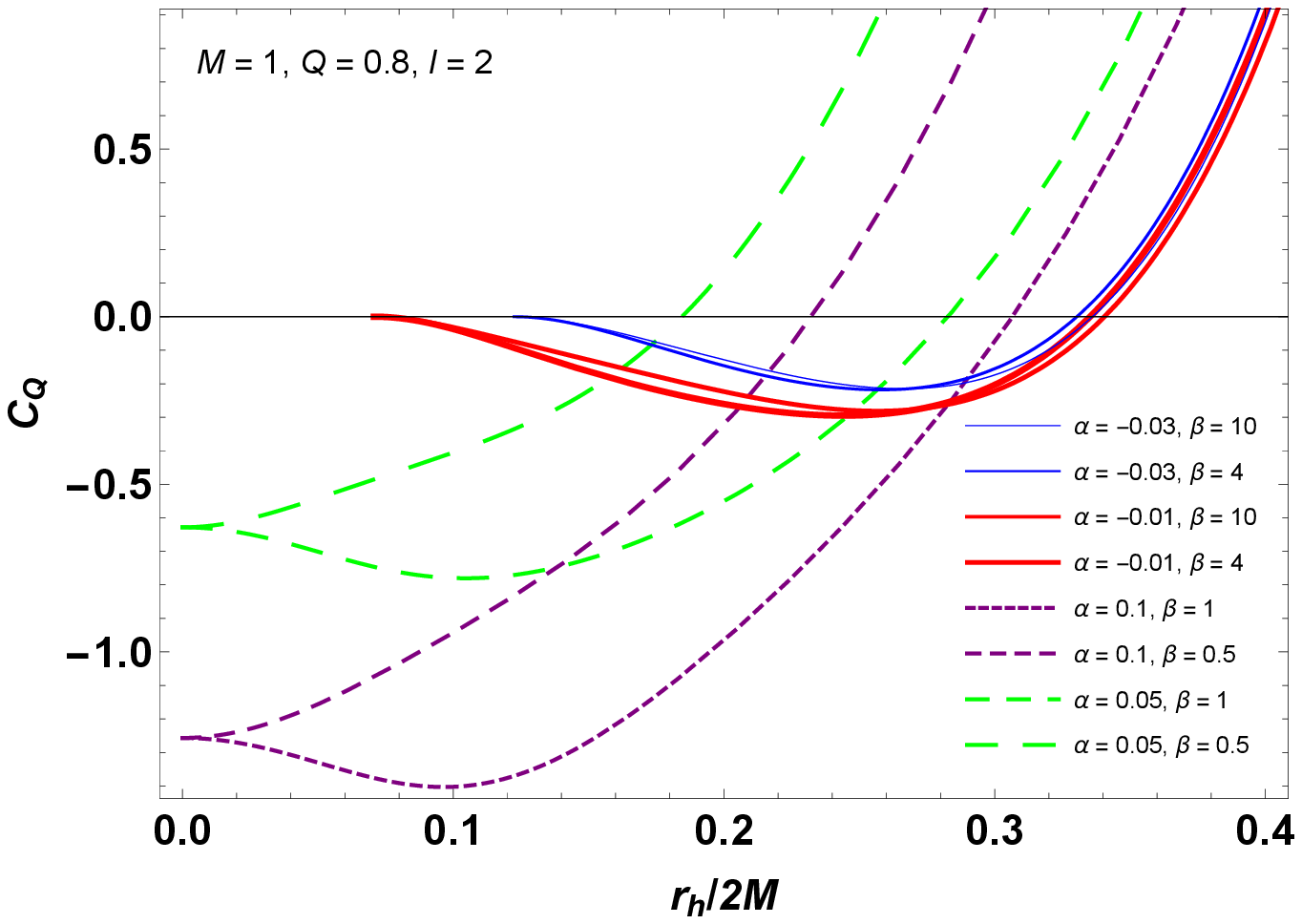}}
\subfigure[]{\label{Sheat_diff_gravities}
\includegraphics[width=4cm,height=3.5cm]{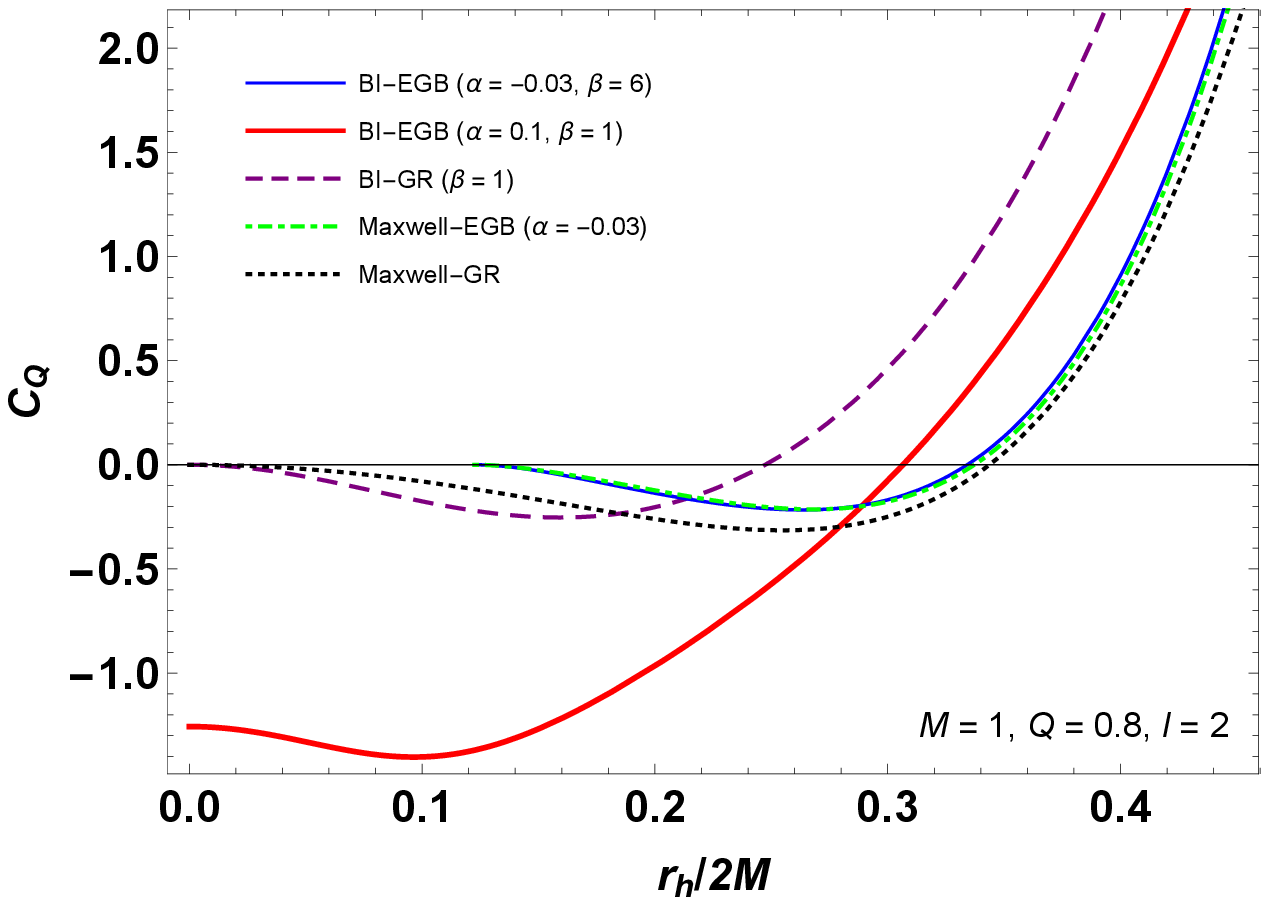}}\\
\caption{The specific heat (a) for some parameter choices and (b) for the BI charged and Maxwell charged solutions in the novel 4D EGB gravity and GR.  } 
\label{specific_heat}
\end{figure}

The specific heat is helpful to analyze the local stability of a black hole solution, and it can be evaluated by
\beq
C_Q=\lt( \fc{\pt M}{\pt T} \rt)_Q=\lt( \fc{\pt M}{\pt r_h} \rt)_Q\lt( \fc{\pt r_h}{\pt T} \rt)_Q.
\eeq
Instead of bothering to list the cumbersome result, we plot the specific heat in Fig.~\ref{specific_heat}. As is shown in Fig.~\ref{Sheat}, it is evident that the black hole has a negative specific heat when the horizon radius is smaller than some critical size $r^s_c$. So only large black holes are stable against fluctuations. A black hole with positive $\alpha$ has larger stable region than that with negative $\alpha$. Moreover, for a given $\alpha$, the smaller the parameter $\beta$, the larger the stable region is. We also show the specific heats for the BI charged and Maxwell charged solutions in the novel 4D EGB gravity and GR in Fig.~\ref{Sheat_diff_gravities}. What they have in common is that large black holes are stable against fluctuations in these theories, since all kinds of black hole solutions shown in Fig.~\ref{Sheat_diff_gravities} are  asymptotic AdS and have the behavior of $a(r) \propto r^2$ at large $r$.

\begin{figure}[htbp]
\centering
\subfigure[$\alpha<0$]{  \label{rsc_EGB_GR_negative}
\includegraphics[width=4cm,height=3.5cm]{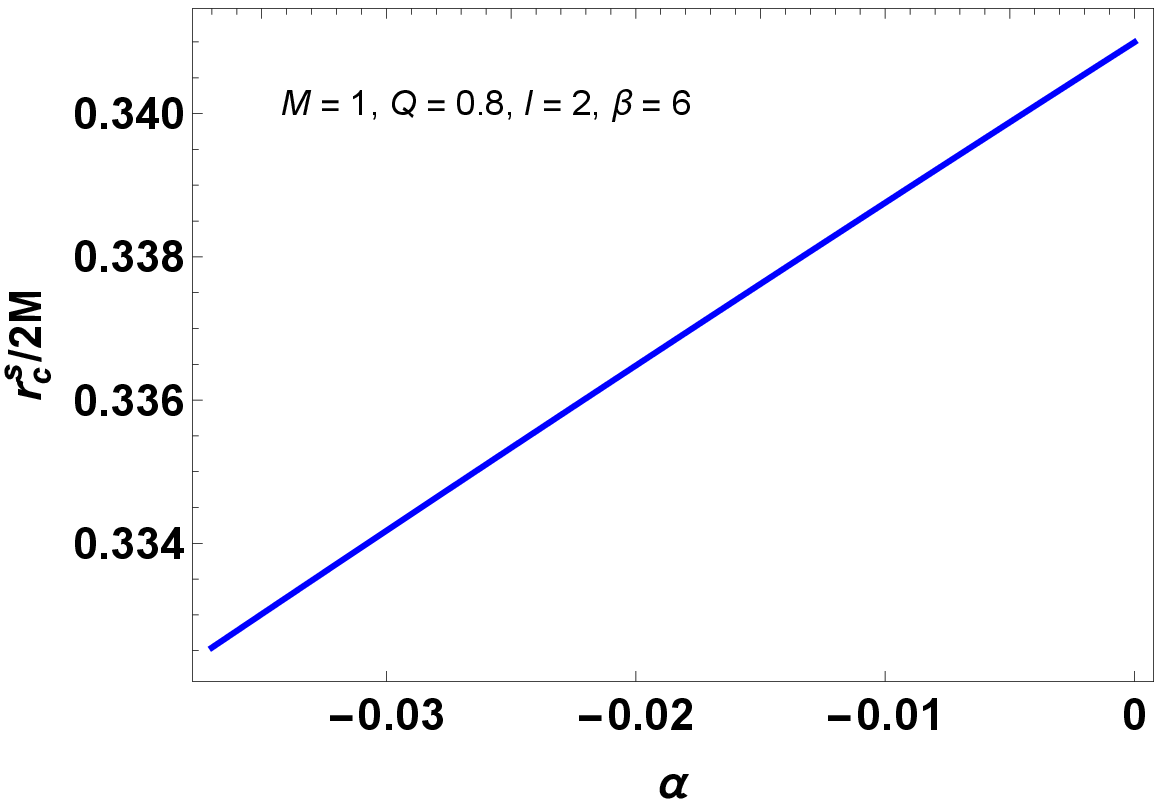}}
\subfigure[$\alpha>0$]{\label{rsc_EGB_GR_positive}
\includegraphics[width=4cm,height=3.5cm]{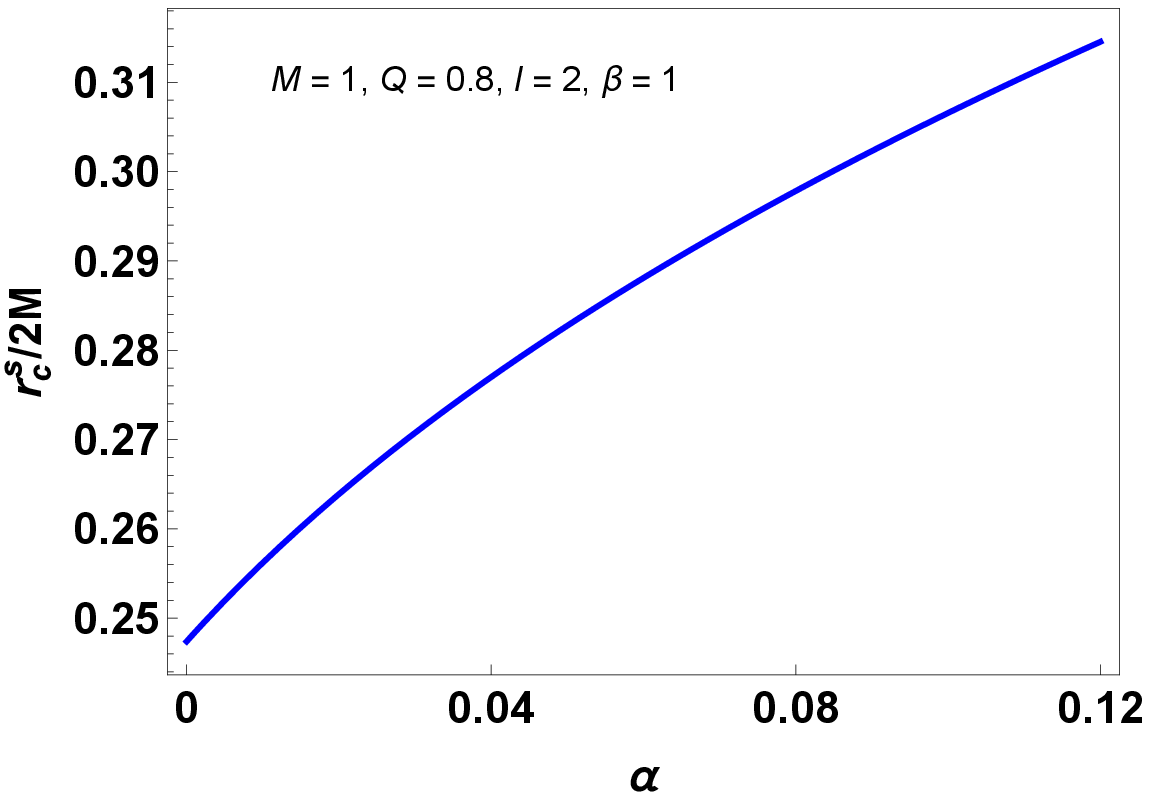}}\\
\caption{The critical size $r^s_c$ of specific heat (a) for negative $\alpha$ and (b) for positive $\alpha$ in the novel 4D EGB gravity. }
\label{rsc_EGB_GR}
\end{figure}

In order to investigate the effect of the GB term on the specific heat, we depict the critical size $r^s_c$ with varying  GB parameter $\alpha$ in Fig.~\ref{rsc_EGB_GR}. When the GB term is switched off, i.e., $\alpha=0$, the BI black hole solution $a(r)$ in the novel 4D EGB gravity will recover the BI black hole solution obtained in EBI theory~\cite{Dey2004}. So as shown in Fig.~\ref{rsc_EGB_GR}, the BI black hole solution with a negative $\alpha$ in the novel 4D EGB gravity has a larger stable region than that in EBI theory, but the solution with a positive $\alpha$ has a smaller region than that in EBI theory.

\begin{figure}[htbp]
\centering
\subfigure[]{  \label{free_energy_dif_para}
\includegraphics[width=4cm,height=3.5cm]{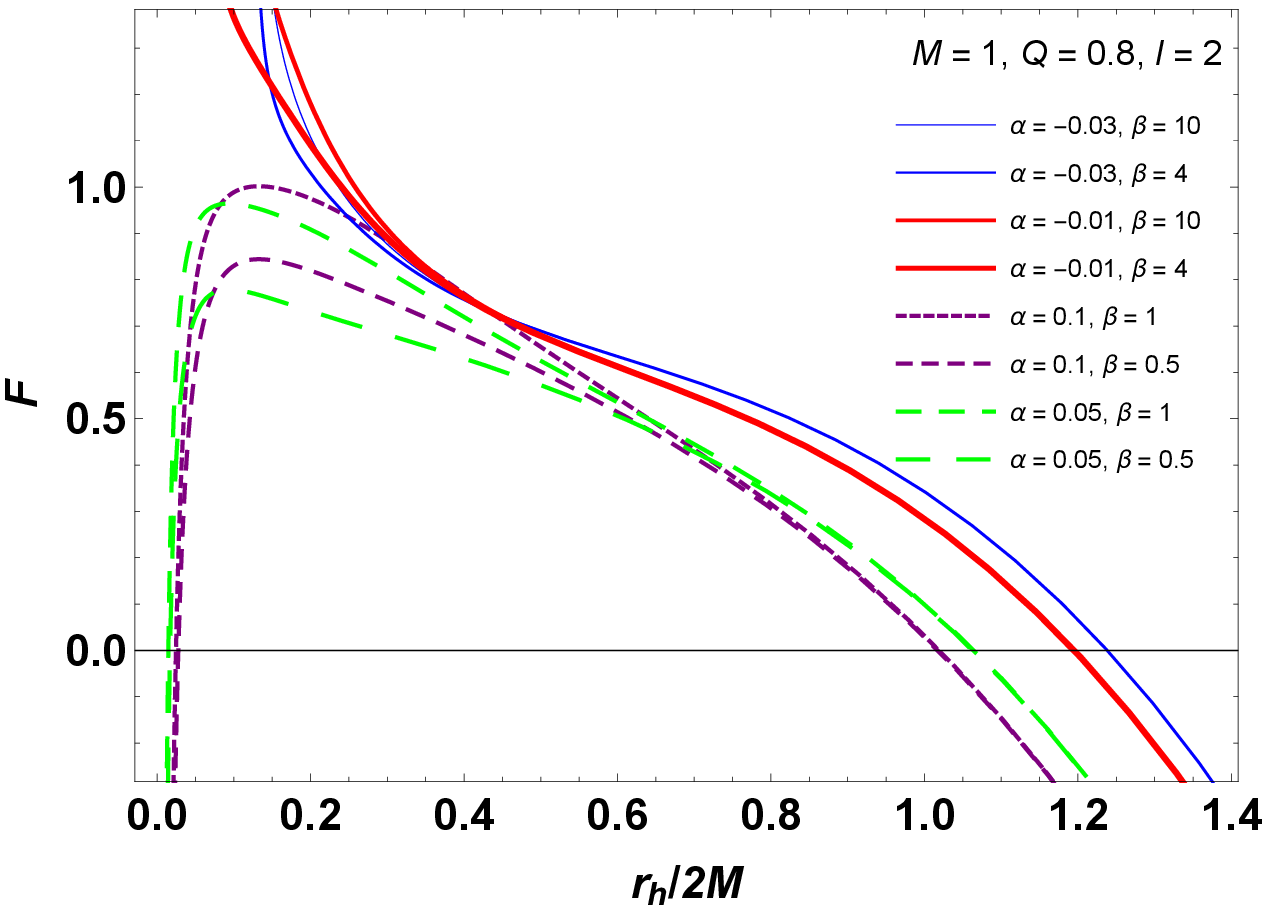}}\subfigure[]{\label{free_energy_dif_gravt}
\includegraphics[width=4cm,height=3.5cm]{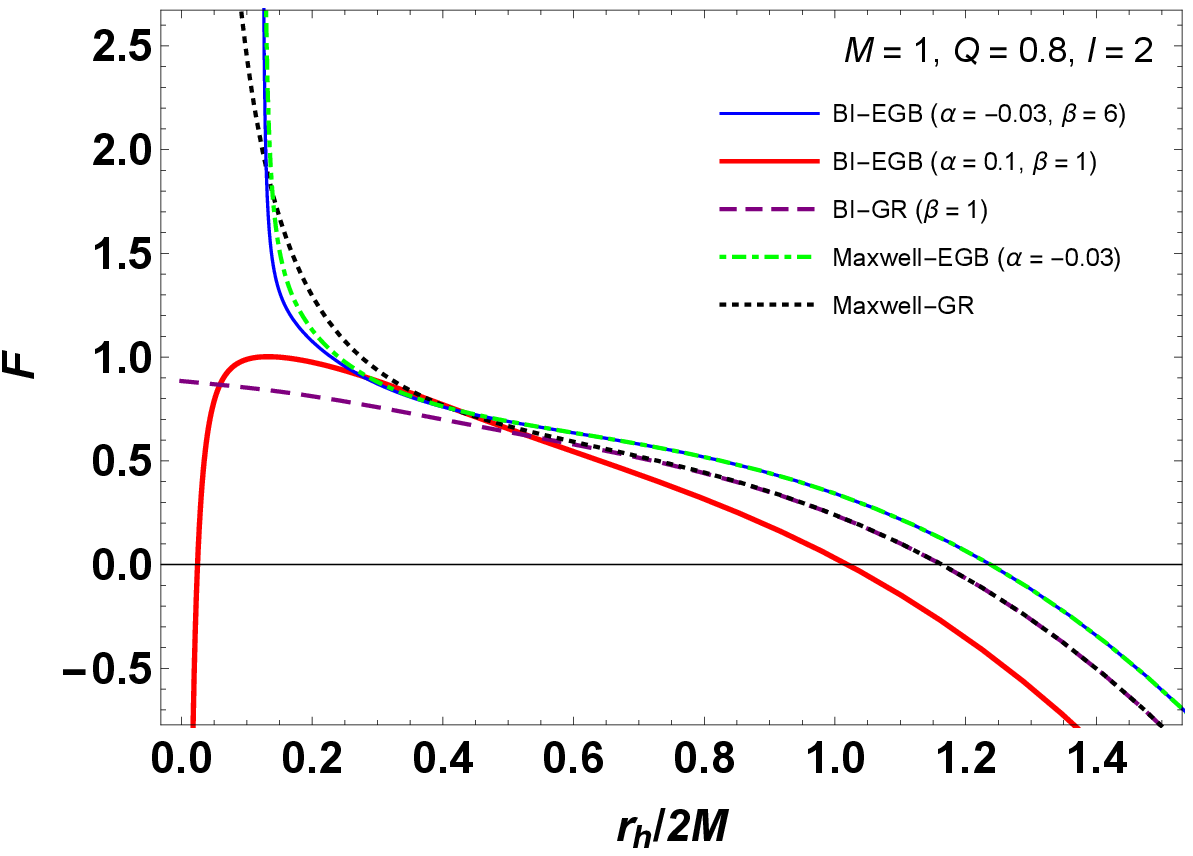}}\\
\caption{The Gibb's free energy (a) for some parameter choices and  (b) for the BI charged and Maxwell charged solutions in the novel 4D EGB gravity and GR.  }
\label{free_energy}
\end{figure}

We further evaluate the Gibb's free energy of the black hole in order to investigate its global stability, which in the canonical ensemble is written as
\beqn
F&=&M-TS\nn\\
&=&\fc{r_h}{2}+\frac{r_h^3}{2 l^2}+\fc{\alpha }{2 r_h}+\fc{\beta ^2 r_h^3}{3}\lt(1-\sqrt{\frac{Q^2}{1+\beta ^2 r_h^4}}\rt)
\nn\\
&&+\frac{2 Q^2}{3 r_h}{~}_2F_1\lt(\fc{1}{4},\fc{1}{2},\fc{5}{4},-\fc{Q^2}{\beta^2 r_{h}^4}\rt)
-\Bigg[\fc{r_h}{2}+\frac{r_h^3}{2 l^2}-\fc{\alpha }{2 r_h}\nn\\
&&+{\beta ^2 r_h^3}\lt(1-\sqrt{1+\frac{Q^2}{\beta ^2 r_h^4}}\rt)\Bigg]\fc{r_h^2+2\alpha\ln\fc{r_h^2}{|\alpha|}}{2(r_h^2+2\alpha)}.
\eeqn

The behavior of the free energy is illustrated in Fig.~\ref{free_energy} for different parameter choices and different black hole solutions. A black hole with a negative free energy is globally stable, but a black hole with a positive one is globally unstable. So as shown in Fig.~\ref{free_energy_dif_para}, for the black hole solution with $\alpha<0$, only when its radius is larger than some critical size $r^f_c$, the black hole is globally stable, while for the solution with $\alpha>0$, both larger and smaller black holes are globally stable.  However, smaller black holes with positive $\alpha$ are locally unstable due to their negative specific heat, as is shown in Fig.~\ref{Sheat}. Moreover, black holes with positive $\alpha$ have larger stable region than that with negative $\alpha$. Further, as shown in Fig. \ref{free_energy_dif_gravt}, BI charged black holes with positive $\alpha$ in the novel 4D EGB gravity have the largest stable region than the others. 

\begin{figure}[htbp]
\centering
\subfigure[$\alpha<0$]{  \label{rfc_EGB_GR_negative}
\includegraphics[width=4cm,height=3.5cm]{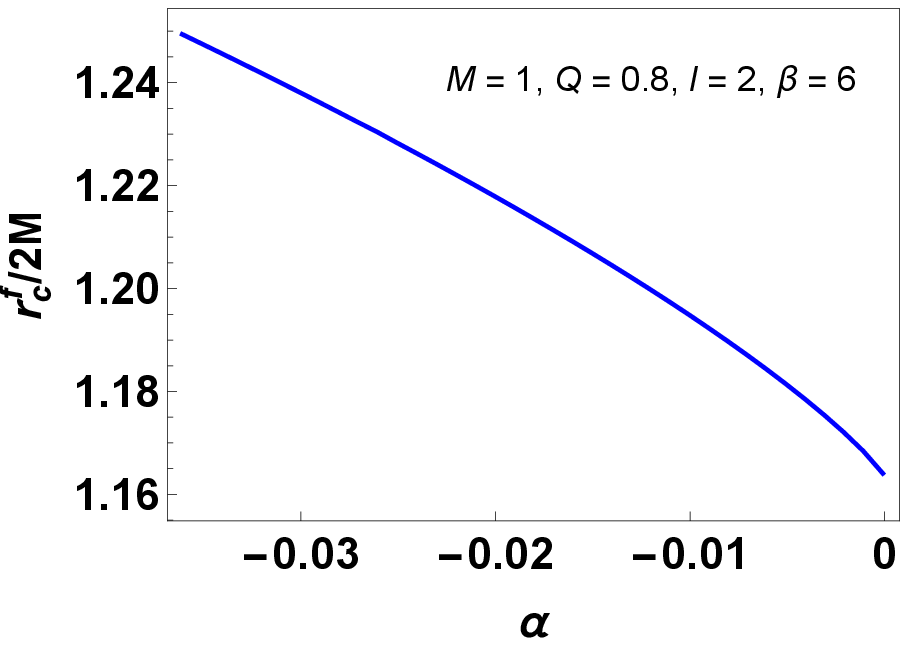}}
\subfigure[$\alpha>0$]{\label{rfc_EGB_GR_positive}
\includegraphics[width=4cm,height=3.5cm]{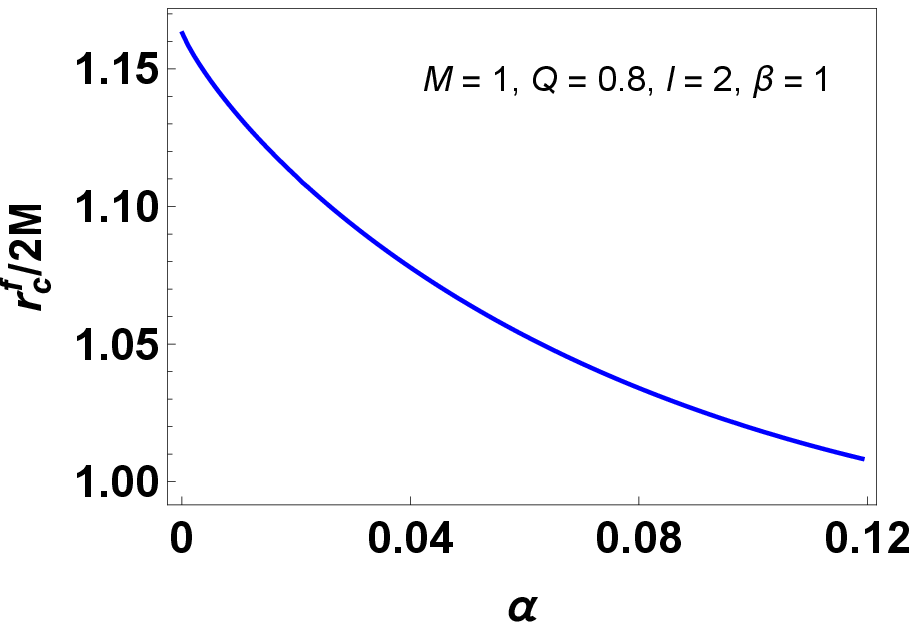}}\\
\caption{The critical size $r^f_c$ of free energy (a) for negative $\alpha$ and (b) for positive $\alpha$ in the novel 4D EGB gravity. }
\label{rfc_EGB_GR}
\end{figure}

In particular, we illustrate the relationship between the critical size $r^f_c$ and the GB parameter $\alpha$ in Fig.~\ref{rfc_EGB_GR}, where a non-vanishing $\alpha$ corresponds to the BI black hole in the novel 4D EGB gravity and a vanishing $\alpha$ corresponds to  the BI black hole in EBI theory. Therefore, it is shown that the BI black hole solution with $\alpha>0$ in the novel 4D EGB gravity has a larger stable region than that in EBI theory, but the solution with $\alpha<0$ has a smaller region than that in EBI theory.

\section{Regain the solution in the regularized 4D EGB gravity}\label{Regain_solution}

In order to get a well defined action principle, H. L\"u  and Y. Pang proposed a procedure for $D \to 4$ limit of EGB gravity \cite{Lu2020}. They started from a $D$-dimensional EGB gravity on a maximally symmetric space of $(D-4)$-dimensions with the metric
\beq
ds^2_D=ds^2_4+e^{2\phi}d\Sigma^2_{D-4,\lambda},
\eeq
where the breathing scalar $\phi$ depends only on the external $4$-dimensional coordinates, $ds^2_4$ is the 4-dimensional line element, and $d\Sigma^2_{D-4,\lambda}$ is the line element of the internal maximally symmetric space with the curvature tensor $R_{abcd}=\lambda(g_{ac} g_{bd}-g_{ad}g_{bc})$. Further, by redefining the GB coupling parameter $\alpha$ as $\alpha\to\fc{\alpha}{D-p}$, and then taking the limit $D \to 4$, they obtained a special scalar-tensor theory that belongs to the family of Horndeski gravity. In this section, we show that the black hole solution (\ref{solution_b}), (\ref{solution_phi}) and (\ref{sol_metric_function}) we obtained in the novel 4D EGB gravity is also the solution of this regularized 4D EGB gravity.

The 4-dimensional regularized gravitational action is given by \cite{Lu2020}
\beqn
\mathcal{S}_\text{reg}&=&\int{}d^4x\sqrt{-g}\bigg[R+\fc{6}{l^2}+\alpha \Big(\phi \mathcal {G} +4 G^{\mu\nu} \pt_\mu \phi\pt_\nu\phi  \nn\\
&&-2\lambda R e^{-2\phi} -4(\pt \phi)^2\Box \phi+2((\pt\phi)^2)^2 \nn\\
&& -12\lambda (\pt\phi)^2 e^{-2\phi}-6\lambda^2 e^{-4\phi} \Big)\bigg].
\eeqn
By including the Lagrangian of the BI electrodynamics (\ref{BI_action}), and substituting the ansatzes for the scalar filed $\phi=\phi(r)$ and the 4D static spherically symmetric metric
\beq
ds^2=-a(r)e^{-2b(r)} dt^2+\frac{dr^2}{a(r)}+r^2(d\theta^2 + \sin\theta^2 d\varphi^2),
\label{metric}
\eeq
 into the action, the effective Lagrangian reads
\beqn
L_\text{eff}&=&e^{-b}\bigg[ 2\lt(1+\fc{3 r^2}{l^2}-a-r a'\rt)+\fc{2}{3}\alpha \phi' \Big( 3r^2 a^2 \phi'^3 \nn\\
\!&&\!-2r f \phi'^2 \lt(4a+r a'-2r a b' \rt)-6a\phi' (1-a-r a'\nn\\
\!&&\!+2r a b') +6(1-a)(a'-2a b') \Big)-6\alpha \lambda^2 r^2 e^{-4\phi}\nn\\
\!&&\! -4\alpha \lambda e^{-2\phi} (1-a -ra'-r^2a'\phi'+2r^2ab' \phi'  \nn\\
\!&&\!+3r^2 a \phi'^2) +4\beta^2 r^{2}\lt(1-\sqrt{1-\beta^{-2 }e^{2b}\Phi'^2} \rt)  \bigg].
\eeqn
The corresponding equations of motion can be obtained by varying the effective acton with respect to $a(r)$, $b(r)$, $\phi(r)$ and $\Phi(r)$. Further, it is easy to see that $b'(r)$ satisfies the equation
\beq
b'\!=\!\frac{ \lt[a \lt(r \phi '-1\rt)^2-\lambda  r^2e^{-2 \phi}  -1\rt] \lt(\phi ''+\phi '^2\rt)}{ \phi ' \left[1-a\left(3+r \phi ' \lt(r \phi '-3\rt)\rt)\rt]+\lambda  r \fc{(r \phi '-1)}{e^{2 \phi} }\!+\!\fc{r}{2 \alpha}}.
\eeq
So following the similar analysis in Ref.~\cite{Lu2020}, $b(r)=0$ is a consistent truncation. In this case,   $a(r)$, $\phi(r)$ and $\Phi(r)$ satisfy the equations
\beqn
1-a-r a'+\fc{2 \alpha  (a-1) a'}{r}-\fc{\alpha  (a-1)^2}{r^2}+\fc{3 r^2}{l^2}\nn\\
+2 \beta ^2 r^2 \lt(1-\sqrt{1+\fc{Q^2}{r^4 \beta ^2}}\rt) &=& 0,\label{EoM2a}\\
1+r^2 \lambda  e^{-2 \phi } -a \left(r \phi '-1\right)^2&=& 0,\label{EoM2b}\\
2 \beta ^2 \Phi'-2 \Phi'^3+r \beta ^2 \Phi'' &=& 0.\label{EoM2c}
\eeqn
It is obviously that $b'(r)=0$ when Eq. (\ref{EoM2b}) holds, hence $b(r)=0$ is indeed a consistent choice. By solving the equations, we have the solution
\begin{flalign}
\Phi(r)&=\fc{Q}{r}{~}_2F_1\lt(\fc{1}{4},\fc{1}{2},\fc{5}{4},-\fc{Q^2}{\beta^2 r^4}\rt),\\
a(r)&=1+\fc{r^2}{2\alpha}\lt\{1\pm\Bigg[1+4\alpha\lt(\fc{2M}{r^3}-\fc{1}{l^2}\nn\rt.\rt.\\
&\lt.\lt.-\fc{2\beta^2}{3}\lt(1-\sqrt{1+\fc{Q^2}{\beta^2 r^4}}\rt)- \fc{4Q}{3r^3}\Phi\rt) \Bigg]^{\fc{1}{2}}\rt\}\\
\phi(r)_\pm&=\ln\fc{r}{L}+\ln\lt[\cosh\psi\pm\sqrt{1+\lambda L^2}\sinh\psi \rt],
\end{flalign}
where $L$ is an arbitrary integration constant and $\psi(r)=\int^r_{r_h}\fc{du}{u\sqrt{a[u]}}$ \cite{Lu2020}. So we regain the black hole solution in this regularized 4D EGB gravity. Note that the electrostatic potential $\Phi(r)$ and the metic function $a(r)$ are independent of the parameter $\lambda$, which is associated with the curvature of the internal maximally symmetric space. 

\section{Conclusions}

In this work, we obtained the BI electric field charged black hole solution in the novel 4D EGB gravity with a negative cosmological constant. In order to have a real black hole solution, the GB coupling parameter $\alpha$ and BI parameter $\beta$ have to be constrained in some regions. It is known that the Maxwell charged solution is real only for negative $\alpha$. While the BI charged black hole solution found in this paper can be real for both positive and negative $\alpha$. Therefore, the BI charged black hole may be superior to the Maxwell charged black hole to be a charged extension of the Schwarzschild-AdS-like black hole in the novel 4D EGB gravity. The black hole has zero, one, or two horizons depending on the parameters. However, since particles incident from infinity can reach the singularity, the black hole solution still suffers the singularity problem. We also explored some simple thermodynamic properties of the BI charged black hole solution. The Smarr formula and the first law of black hole thermodynamics were verified. By evaluating the specific heat and Gibb's free energy, we showed that the black hole is thermodynamically stable when the horizon radius is large, but is unstable when it is small. This is a well-known property of the AdS black holes. Moreover, a black hole with positive $\alpha$ has larger stable region than that with negative $\alpha$. At last, we also regained the black hole solution in the regularized 4D EGB gravity proposed by H. L\"u  and Y. Pang.

Black hole thermodynamics in AdS space has been of great interest since it possesses some interesting phase transitions and critical phenomena as seen in normal thermodynamic systems. Further thermodynamic properties of the BI charged black hole in the novel 4D EGB gravity is left for our future investigations.

\section*{ACKNOWLEDGMENTS}

We would like to thank Yu-Peng Zhang for helpful discussion.
K. Yang acknowledges the support of ``Fundamental Research Funds for the Central Universities" under Grant No. XDJK2019C051. This work was supported by the National Natural Science Foundation of China (Grants No. 11675064, No. 11875151 and No. 11947025).


\end{document}